\documentclass[conference]{IEEEtran}
\IEEEoverridecommandlockouts
\usepackage{cite}
\usepackage{amsmath,pifont,amssymb,amsfonts}
\usepackage{algorithmic}
\usepackage{graphicx}
\usepackage{textcomp}
\usepackage{xcolor}
\usepackage{balance}
\usepackage{dblfloatfix}
\usepackage{tcolorbox}
\usepackage{url}
\usepackage[hidelinks]{hyperref}

\def\BibTeX{{\rm B\kern-.05em{\sc i\kern-.025em b}\kern-.08em
    T\kern-.1667em\lower.7ex\hbox{E}\kern-.125emX}}
\begin{document}

\title{RAILS: Retrieval-Augmented Intelligence for Learning Software Development
}

\author{\IEEEauthorblockN{Wali Mohammad Abdullah}
\IEEEauthorblockA{\textit{Mathematics \& Information Technology} \\
\textit{Concordia University of Edmonton}\\
Edmonton, Alberta, Canada \\
wali.abdullah@concordia.ab.ca}
\and
\IEEEauthorblockN{Md. Morshedul Islam}
\IEEEauthorblockA{\textit{Mathematics \& Information Technology} \\
\textit{Concordia University of Edmonton}\\
Edmonton, Alberta, Canada \\
mdmorshedul.islam@concordia.ab.ca}
\and
\IEEEauthorblockN{Devraj Parmar}
\IEEEauthorblockA{\textit{Mathematics \& Information Technology} \\
\textit{Concordia University of Edmonton}\\
Edmonton, Alberta, Canada \\
dvparmar@student.concordia.ab.ca}
\and
\IEEEauthorblockN{Happy Hasmukhbhai Patel}
\IEEEauthorblockA{\textit{Mathematics \& Information Technology} \\
\textit{Concordia University of Edmonton}\\
Edmonton, Alberta, Canada \\
hpatel13@student.concordia.ab.ca}
\and
\IEEEauthorblockN{Sindhuja Prabhakaran}
\IEEEauthorblockA{\textit{Mathematics \& Information Technology} \\
\textit{Concordia University of Edmonton}\\
Edmonton, Alberta, Canada \\
sprabhak@student.concordia.ab.ca}
\and
\IEEEauthorblockN{Baidya Saha}
\IEEEauthorblockA{\textit{Mathematics \& Information Technology} \\
\textit{Concordia University of Edmonton}\\
Edmonton, Alberta, Canada \\
baidya.saha@concordia.ab.ca}
}

\maketitle

\begin{abstract}
Large Language Models (LLMs) like GPT-3.5-Turbo are increasingly used to assist software development, yet they often produce incomplete code or incorrect imports, especially when lacking access to external or project-specific documentation. We introduce RAILS (Retrieval-Augmented Intelligence for Learning Software Development), a framework that augments LLM prompts with semantically retrieved context from curated Java resources using FAISS and OpenAI embeddings. RAILS incorporates an iterative validation loop guided by compiler feedback to refine suggestions. We evaluated RAILS on 78 real-world Java import error cases spanning standard libraries, GUI APIs, external tools, and custom utilities. Despite using the same LLM, RAILS outperforms baseline prompting by preserving intent, avoiding hallucinations, and surfacing correct imports even when libraries are unavailable locally. Future work will integrate symbolic filtering via PostgreSQL and extend support to other languages and IDEs.
\end{abstract}


\section{Introduction}

Large Language Models (LLMs) such as GPT-3.5 and GPT-4 are increasingly being integrated into software development workflows due to their ability to generate, complete, and repair source code across programming languages \cite{kabir2025zs4c, wang2021codet5}. Despite these capabilities, LLMs still exhibit notable limitations, particularly hallucinated outputs, lack of awareness of project-specific context, and the generation of outdated or incorrect API usage \cite{zhang2019analyzing, zhou2022docprompting}. These limitations are especially critical in Java, where missing or incorrect import statements frequently cause compilation errors that are not always easy to debug or resolve automatically \cite{zhang2021study}.

In this work, we define a baseline model as GPT-3.5-Turbo used in a zero-shot setting, without any external retrieval or augmentation—i.e., prompted directly with incomplete or erroneous code. This reflects how LLMs are typically used in practice and serves as our primary baseline for comparing against RAILS. We refer to this setup as baseline prompting throughout the paper.

We introduce RAILS (Retrieval-Augmented Intelligence for Learning Software Development), a prototype framework that enhances LLM-based code generation using retrieval-augmented generation (RAG) to address these shortcomings. RAILS performs semantic search over a curated corpus of Java documentation and tutorials using a vector database and integrates relevant content directly into the prompt. Additionally, a validation loop re-invokes the model based on compiler diagnostics, enabling multi-turn refinement of broken code. This reduces hallucinations and surfaces context-aware corrections, particularly beneficial for developers who may be unfamiliar with certain imports or third-party libraries.

Our current work targets Java import-related issues, a known high-frequency source of developer frustration and build failure on platforms like Stack Overflow \cite{zhang2019analyzing, xia2017developers, phan2018statistical, zhang2021study}. While the focus is on Java, RAILS is designed to be language-agnostic and extensible to complex code ecosystems—including those found in high performance computing (HPC) environments. Its retrieval-feedback loop can generalize to scenarios such as missing modules in MPI/C++ workflows, OpenMP pragma misuse, and incompatible dependency chains in scientific software stacks. Future extensions include hybrid symbolic-semantic retrieval (e.g., PostgreSQL with \texttt{pgvector}), runtime validation, and IDE integration (e.g., Visual Studio Code), paving the way for real-time, verifiable developer assistance across academic and enterprise settings.

\textbf{Contributions.} This paper makes the following key contributions:
\begin{itemize}
\item We propose RAILS, a retrieval-augmented framework that addresses a widespread but underexplored challenge in LLM-assisted software development: import-related compilation errors.
\item We design a semantic retrieval and prompt construction pipeline that augments LLM responses with contextual documentation and integrates a validation feedback loop driven by compiler diagnostics.
\item We evaluate RAILS across 78 Java import-related scenarios—including standard APIs, third-party libraries, GUIs, and custom utilities—and show that it consistently outperforms baseline prompting in semantic accuracy, compilation success, and intent preservation.
\end{itemize}

\begin{tcolorbox}[colback=gray!10, colframe=black!50, title=\textbf{Why Use RAILS Instead of Existing Tools?}]
While proprietary tools like GitHub Copilot or ChatGPT plugins provide AI-powered code assistance, they often operate as black boxes—offering no control over what context is retrieved or how it is used. In contrast, RAILS offers a transparent and modular retrieval-augmented pipeline where every stage—from semantic search to compiler-informed refinement—is explicitly inspectable, configurable, and reproducible.

RAILS supports model-agnostic deployment (OpenAI or open-weight models), and allows retrieval from private or custom documentation. Its compiler feedback loop mimics real-world debugging behavior and is absent in most commercial tools. This makes RAILS well-suited for research, education, and integration into trusted software development workflows where interpretability, cost control, and fine-grained control are essential.
\end{tcolorbox}

We organize the rest of this paper as follows: Section~\ref{sec:architecture} presents the architecture of the RAILS framework, detailing its semantic retrieval, prompt construction, and validation loop. Section~\ref{sec:evaluation} describes the experimental setup, case selection, and benchmarking environments. Section~\ref{sec:results} summarizes the empirical findings, while Section~\ref{sec:discussion} discusses the broader implications and potential for future extensions. Finally, Section~\ref{sec:conclusion} concludes the paper and outlines directions for ongoing and future work.

\section{System Architecture}\label{sec:architecture}

\begin{figure*}[!ht]
  \centering
  \includegraphics[width=0.785\textwidth]{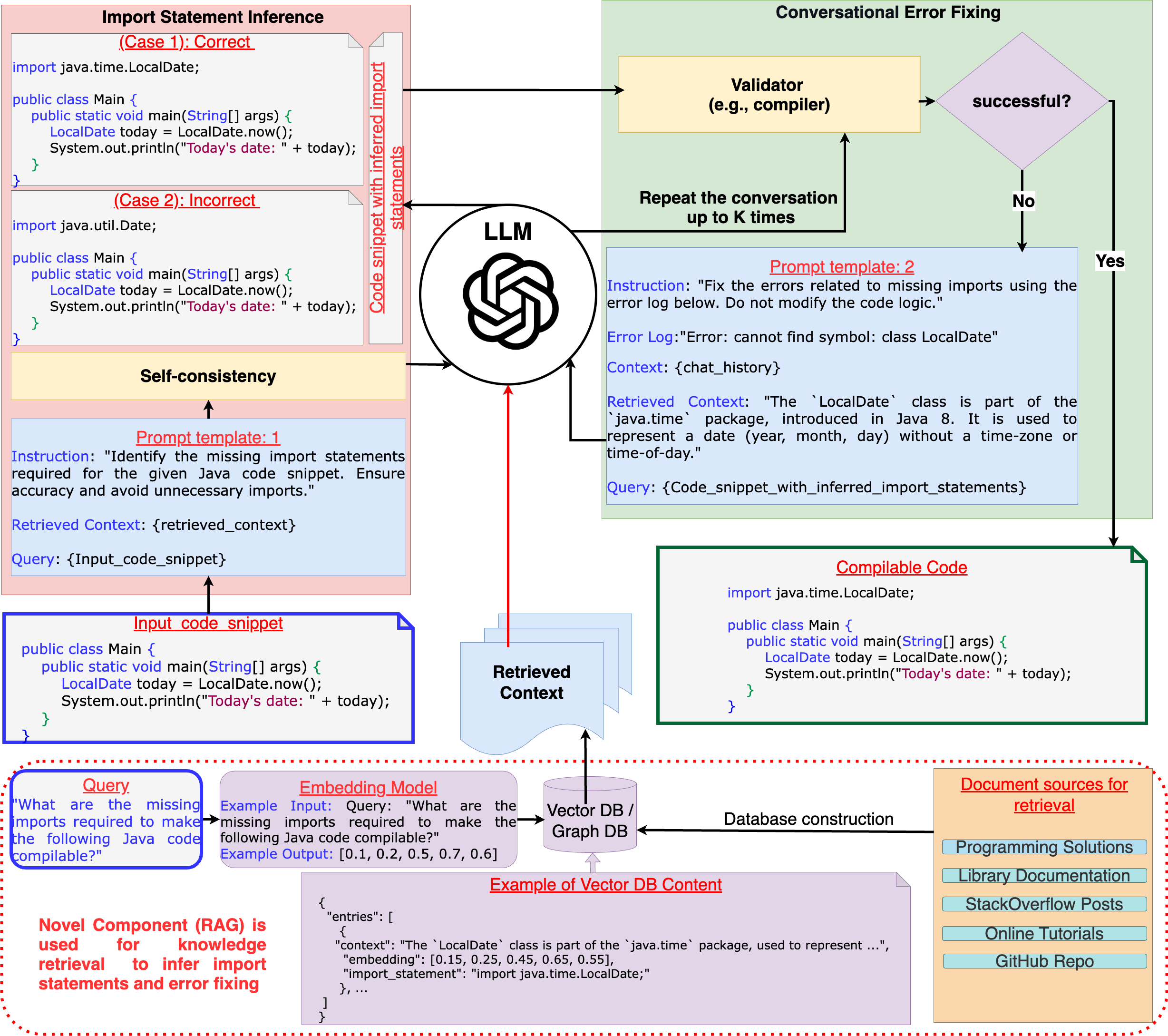}
  \caption{RAILS methodology: semantic retrieval, prompt generation, and validation feedback loop for context-aware code repair.}
  \label{fig:rails_methodology}
\end{figure*}

RAILS is designed to enhance the accuracy of LLM-based code generation by integrating semantic retrieval and iterative validation. As illustrated in Figure~\ref{fig:rails_methodology}, the architecture is composed of three key modules: a semantic retriever, a prompt construction engine, and a validation feedback loop.

\begin{itemize}
    \item \textbf{Semantic Retriever:} Uses OpenAI embeddings and FAISS~\cite{johnson2017faiss} to perform dense vector search over Java documentation and tutorials. This module identifies semantically similar code snippets or explanations relevant to the current coding error.
    \item \textbf{Prompt Construction:} Builds an augmented prompt combining the original code, retrieved context, compiler diagnostics, and previous LLM outputs. Prompting is managed using LangChain~\cite{zhou2022docprompting}, enabling multi-turn refinement.
    \item \textbf{Validation Loop:} Compiles the generated code and feeds any resulting errors back into the next prompt iteration. This loop mimics real-world debugging and helps guide the LLM toward a valid and contextually solution.
\end{itemize}

\subsection{Vector Database and Retrieval}

To enable semantic retrieval, RAILS integrates a vector store powered by FAISS~\cite{johnson2017faiss}, built from a curated corpus of Java documentation and tutorial content, including official Oracle guides~\cite{jdk24tutorial}. The corpus is segmented using LangChain’s \texttt{RecursiveCharacterTextSplitter}, which splits documents into overlapping chunks of 300 characters with a 50-character overlap—parameters chosen to preserve context while enhancing retrieval precision.

Each chunk is embedded using OpenAI’s \texttt{text-embedding-ada-002} model via the LangChain \texttt{OpenAIEmbeddings} interface. These embeddings are indexed and persisted using FAISS. The vector index is stored locally using the \texttt{save\_local} method and later accessed via LangChain’s retriever API. This setup allows RAILS to query the vector store at runtime and inject the most relevant knowledge chunks into the prompt for error correction.

\begin{figure}[ht]
  \centering
  \includegraphics[width=0.47\textwidth]{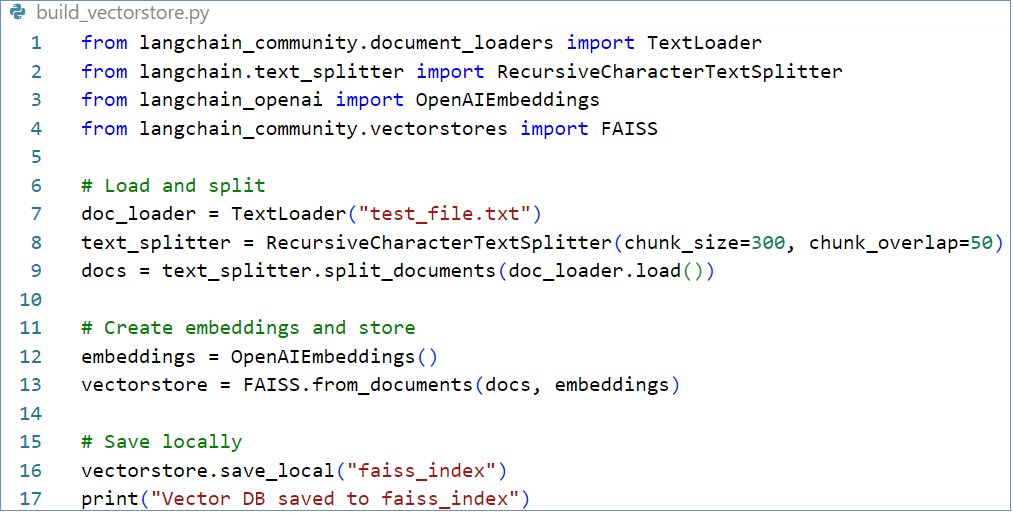}
  \caption{Code snippet for building the vector store from Java documentation using FAISS and OpenAI embeddings.}
  \label{fig:rag_code}
\end{figure}

The end-to-end data preparation and vector store creation workflow is illustrated in Figure~\ref{fig:rag_code}.

While FAISS enables low-latency vector similarity search, it lacks symbolic querying capabilities and can struggle with large documentation sets due to memory constraints. To address this, we plan to migrate to a hybrid retrieval backend using PostgreSQL and the \texttt{pgvector} extension. This would allow symbolic filters (e.g., keyword or library type) to work alongside semantic similarity, enabling more precise and scalable retrieval pipelines within IDEs. This approach follows recent findings advocating hybrid retrieval for reliable system behavior~\cite{goyal2022retrieval, hassan2024rethinking}.

\subsection{Prompt Construction}

\begin{figure}[ht]
  \centering
  \includegraphics[width=0.47\textwidth]{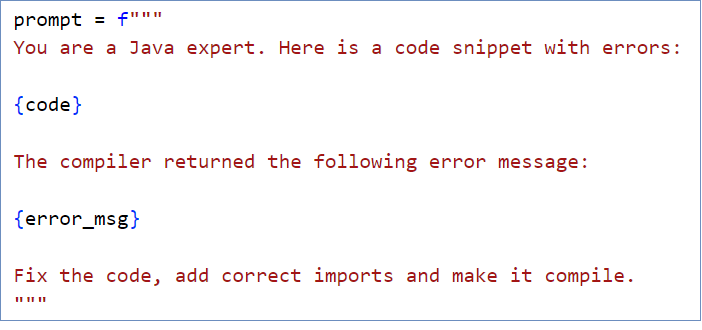}
  \caption{Baseline prompt without retrieval context.}
  \label{fig:vanilla_prompt}
\end{figure}

RAILS and the base prompting approach differ significantly in how they construct instructions for the LLM to generate correct code. The baseline model uses a simple prompt that only includes the broken Java code and the associated compiler error, relying entirely on the model’s internal training. This often leads to hallucinated fixes, especially in the absence of contextual knowledge~\cite{zhang2019analyzing, zhou2022docprompting}. An example of this baseline prompt is shown in Figure~\ref{fig:vanilla_prompt}.

\begin{figure}[ht]
  \centering
  \includegraphics[width=0.47\textwidth]{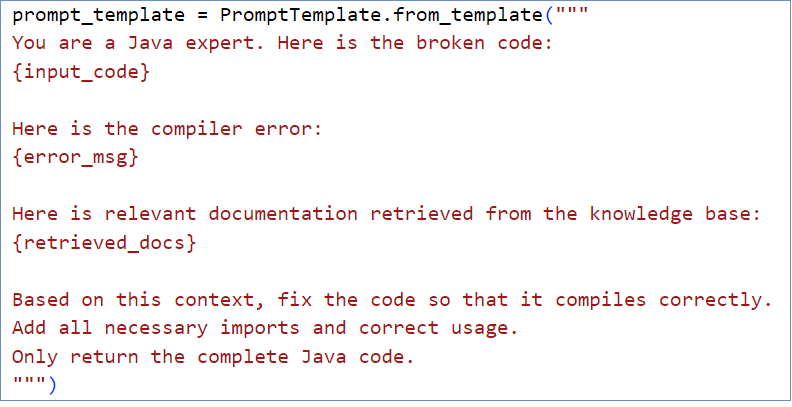}
  \caption{RAILS prompt enriched with retrieved documentation and error feedback.}
  \label{fig:rails_prompt}
\end{figure}

However, RAILS adopts a retrieval-augmented strategy. It enriches the prompt with semantically retrieved documentation relevant to the faulty code and error messages. This additional context is pulled from a FAISS-based vector database built from Java tutorials and official documentation~\cite{jdk24tutorial, johnson2017faiss, gu2021deep, husain2019codesearchnet}. The resulting prompt includes four elements: the broken code, the compiler error, the retrieved context, and an instruction to generate compilable output. Figure~\ref{fig:rails_prompt} illustrates an example of the RAILS prompt format.

Furthermore, RAILS supports iterative refinement. If a generated fix still fails to compile, the feedback loop feeds the new error message into the next prompt. This multi-turn repair mechanism follows principles proposed in~\cite{lu2021fantastically, zhang2023steam, zhou2022large}, and supports better adaptation to specific APIs~\cite{saifullah2019learning}.

While tools like GitHub Copilot also assist with import errors, they offer limited transparency and do not expose retrieval mechanisms or compiler feedback loops. RAILS, in contrast, provides full control over retrieval sources, prompt construction, and validation steps, making it more suitable for IDE integration, debugging, and academic reproducibility.

\section{Empirical Evaluation}\label{sec:evaluation}

To assess the effectiveness of RAILS compared to baseline prompting with LLMs, we designed an evaluation framework targeting a recurring class of Java programming challenges: import-related compilation errors. We curated a total of 78 Java code snippets reflecting real-world issues frequently encountered on developer forums such as Stack Overflow. These examples span standard library omissions (e.g., \texttt{java.util.ArrayList}), deprecated API usage (e.g., \texttt{java.util.Date}), graphical components (e.g., Swing and JavaFX), external dependencies (e.g., Apache Commons, Gson), and references to user-defined utilities.

Each code sample was evaluated under two prompting strategies—baseline prompting and the RAILS framework—with compilation-based validation to assess correctness and semantic alignment. The average end-to-end latency of a RAILS iteration—including retrieval, prompt construction, and LLM inference—was approximately 4.1 seconds per case on our local machine. In contrast, baseline prompting without retrieval averaged 3.3 seconds per case. The additional latency in RAILS is primarily due to the semantic retrieval step, which accounts for roughly 20–30\% of total runtime. This trade-off is offset by RAILS’s higher semantic accuracy and is expected to reduce with asynchronous retrieval in future IDE integrations. Portability was confirmed with consistent behavior on Google Colab, supporting reproducibility across environments.

\subsection{Evaluation Setup}

Experiments were conducted in two computational environments to validate performance across different hardware configurations:

\begin{itemize}
    \item \textbf{Local Machine:} A Windows Subsystem for Linux (WSL) environment on a laptop equipped with an Intel(R) Core(TM) i7-8650U CPU @ 1.90GHz, 16 GB RAM, and Ubuntu 24.04.
    \item \textbf{Google Colab (Cloud Environment):} A cloud-hosted runtime with 12.67 GB RAM and Python 3 running on a Google Compute Engine backend. This setup validated reproducibility and generalizability of RAILS in an online, server-based environment.
\end{itemize}

Each case was processed using two pipelines:

\begin{enumerate}
    \item \textbf{Baseline Prompting:} GPT-3.5-turbo was prompted with the broken Java snippet and corresponding compilation error, without access to any external documentation.
    \item \textbf{RAILS (RAG-Augmented Prompting):} GPT-3.5-turbo received the same inputs along with semantically retrieved context drawn from a FAISS vector store built from Java tutorials, API documentation, and community-contributed examples \cite{jdk24tutorial}.
\end{enumerate}

\subsection{Experiment Pipeline}

For each of the 78 code examples, both prompting approaches followed a consistent three-step evaluation loop:

\begin{enumerate}
    \item The original code was compiled and error messages were logged.
    \item The LLM (with or without RAG) generated a fix based on the code and error.
    \item The proposed fix was recompiled, and the outcome—success, partial fix, or failure—was recorded.
\end{enumerate}

Each trial produced logs capturing the LLM output, compiler responses, and (for RAILS) the retrieved documents. These logs were stored in separate files for both approaches, \texttt{log\_baseline.txt} and \texttt{log\_rails.txt}, to ensure consistency and traceability. We validated semantic correctness for non-compiling cases by resolving missing dependencies and confirming expected runtime behavior. Although most RAILS fixes succeeded in the first attempt, the feedback loop remains critical for future multi-turn or cross-file repair scenarios.

\subsection{Reproducibility Details}

The RAILS vector store was built using content from the official Java documentation \cite{jdk24tutorial}. The documents were preprocessed and segmented using LangChain’s \texttt{RecursiveCharacterTextSplitter}, producing overlapping text chunks of 300 characters with a 50-character overlap. We embedded 1,456 chunks using OpenAI's \texttt{text-embedding-ada-002} model (API v2.1, temperature=0), resulting in a FAISS index of approximately 22MB.

During inference, the retriever used cosine similarity to return the top $k=4$ most relevant chunks per query. These chunks were injected into the LLM prompt for context-aware code generation. 
To facilitate replication and transparency, all code, data, and logs associated with RAILS are available online at~\cite{rails2025}.

\section{Results}\label{sec:results}

We observed notable performance differences between RAILS and baseline prompting, particularly in cases involving external libraries, GUI frameworks, and project-specific utilities.

\begin{table}[!htb]
\centering
\caption{Semantic Correctness of Code Generated by RAILS vs. Baseline Prompting}
\begin{tabular}{|l|c|c|}
\hline
\textbf{Case Type} & \textbf{RAILS} & \textbf{Baseline Prompting} \\
\hline
Standard JDK Import (e.g., & \checkmark & \checkmark \\
\texttt{ArrayList}, \texttt{HashMap}) & & \\ \hline
Deprecated API Fix & \checkmark & \checkmark \\
(\texttt{java.util.Date} $\rightarrow$ & & \\
\texttt{LocalDate}) & & \\ \hline
Swing UI Component & \checkmark & \checkmark \\
(\texttt{JFrame}) & & \\ \hline
Java.nio.file Usage & \checkmark & \checkmark \\
(\texttt{Paths}, \texttt{Files}) & & \\ \hline
External Lib & \checkmark & \ding{55} (Missing import) \\
(Apache Commons I/O) & & \\ \hline
External Lib (Gson, Text) & \checkmark & \ding{55} \\
& & (No retrieval or awareness)\\ \hline
GUI with JavaFX & \checkmark & \ding{55} (Incorrect or failed usage) \\ 
(\texttt{Alert}) & & \\ \hline
Custom Utility Class & \checkmark & \ding{55} (Hallucinated \\
(e.g., \texttt{MySpecialUtils}) & & or substituted logic) \\
\hline
\end{tabular}
\label{tab:comparison_summary}
\end{table}

As shown in Table~\ref{tab:comparison_summary}, both approaches performed reliably on standard Java SE constructs such as collections, date/time handling, and file path utilities. However, a performance gap emerges in domains that require contextual or external knowledge, including external libraries (e.g., Apache Commons), JavaFX, and custom utilities.

In all 78 cases, RAILS maintained 100\% semantic correctness in its proposed solutions. While 18 examples failed to compile due to missing runtime dependencies (e.g., unavailable JARs), the retrieved context and generated logic were correct. In contrast, the baseline model failed to compile 14 cases and produced semantically incorrect outputs in four others, often rewriting user-defined methods or substituting custom code with standard library approximations.

\vspace{1cm}
\textbf{Observations:}
\begin{itemize}
    \item Both pipelines succeeded for standard Java SE examples including core classes like \texttt{ArrayList}, \texttt{HashMap}, and \texttt{Paths}.
    \item RAILS retrieved and integrated context for external dependencies (e.g., Commons I/O, Gson) even without having access to the libraries in the local Java environment.
    \item In GUI (JavaFX) and custom utility cases, RAILS preserved semantic intent; the baseline approach frequently substituted logic or fabricated fallback code.
    \item Hallucinations in the baseline pipeline were observed in four cases (e.g., replacing \texttt{MySpecialUtils.normalize()} with unrelated built-ins), highlighting a lack of retrieval-grounding.
\end{itemize}

\section{Discussion}\label{sec:discussion}

The expanded evaluation in Section~\ref{sec:results} demonstrates that RAILS consistently produces semantically correct and intent-preserving code fixes, even in scenarios involving deprecated APIs, third-party libraries, JavaFX GUI components, and user-defined utilities. These improvements stem from RAILS’s retrieval-augmented prompting strategy, which supplements LLM reasoning with documentation-grounded context retrieved at runtime.

In contrast, baseline prompting frequently struggled to resolve import-related issues that fell outside the standard Java SE domain. Notably, the baseline model hallucinated replacement logic in four cases and failed to compile in fourteen more—typically due to missing external packages or incorrect import suggestions. RAILS, by comparison, failed to compile in 18 cases but maintained semantic correctness in all of them by issuing logically valid and well-structured code. This highlights the value of retrieval grounding in reducing semantic drift~\cite{zhou2022docprompting, zhou2022large, saifullah2019learning}.

\begin{figure}[ht]
  \centering
  \includegraphics[width=0.45\textwidth]{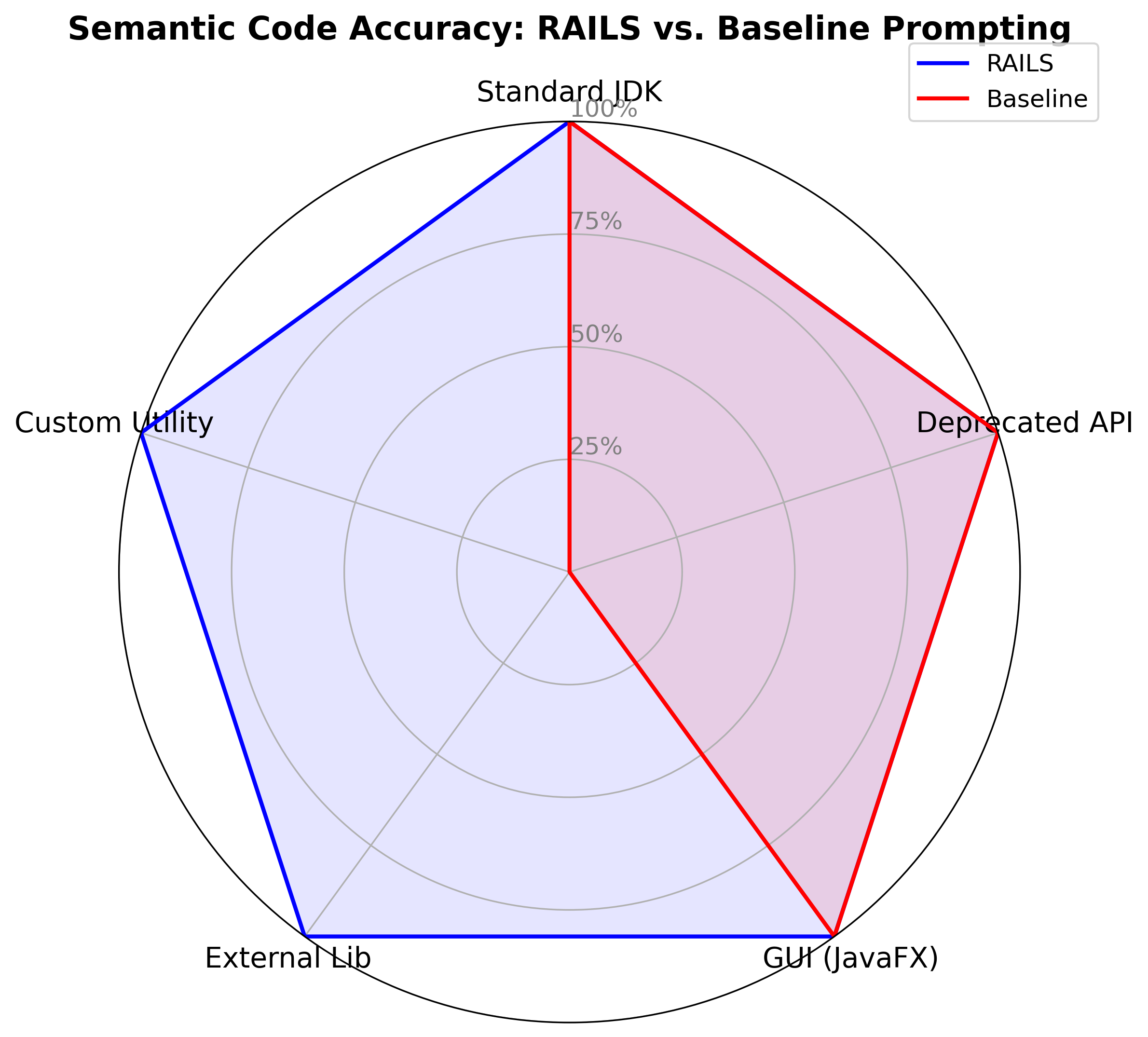}
  \caption{Radar plot showing semantic code accuracy across categories. RAILS (blue) sustains high correctness, while Baseline (red) underperforms in complex import scenarios.}
  \label{fig:deficiency_plot}
\end{figure}

To further illustrate these trends, Figure~\ref{fig:deficiency_plot} presents a radar chart comparing semantic correctness across five representative import-related error categories. RAILS (blue) maintains near-perfect coverage in all categories, including external libraries and custom project code. Baseline prompting (red), however, underperforms in the JavaFX and external library domains and demonstrates zero correctness in custom utility scenarios—primarily due to hallucinations or logic substitutions.

This consistency makes RAILS a strong candidate for real-world adoption. As modern software increasingly integrates external APIs, community libraries, and modular internal codebases, developer tools must handle ambiguous or context-dependent imports. Retrieval-augmented repair frameworks like RAILS can bridge this gap by offering grounded, verifiable, and reproducible solutions. In team-based environments or during onboarding, such systems improve developer confidence and reduce debugging overhead.

% However, limitations remain. RAILS relies on a static, pre-indexed corpus that may become outdated as libraries evolve. The prototype is also evaluated only for Java, and scaling to other languages will require new corpora, updated embedding pipelines, and potentially language-specific prompt tuning. Nonetheless, RAILS provides a modular and extensible framework to build on—particularly as offline or open-weight LLMs improve and symbolic-semantic retrieval strategies (e.g., PostgreSQL \texttt{pgvector} + BM25) become more accessible.

\section{Design Scope and Future Work}

\subsection{Current Scope and Model Constraints}
RAILS is currently scoped to Java import repair, where semantic evaluation and compilation validation can be reliably tested. While our prototype uses GPT-3.5-turbo, RAILS supports interchangeable LLMs, including open-weight and cost-effective alternatives. Token window constraints may affect long or multi-file inputs, motivating future integration of summarization and hierarchical prompting.

\subsection{Retrieval Precision}
Our current dense retrieval setup with FAISS enables low-latency matching but may miss certain syntactic cues. We plan to adopt hybrid retrieval via PostgreSQL and \texttt{pgvector}, enabling symbolic filters (e.g., by library or API) to complement semantic similarity for greater precision.

\subsection{Extensibility Across Languages and Environments}
RAILS’s modular design is language-agnostic and intended for broader repair tasks. Planned extensions include support for C++ and Python (e.g., OpenMP, pip imports), runtime bug repair, and plugin integration with IDEs such as VS Code. These directions aim to support real-time, context-aware developer assistance across diverse software ecosystems.

\subsection{Evaluation Comparisons}
Future evaluations will include comparisons against retrieval-only pipelines (e.g., BM25), static analysis tools, and fine-tuned small models. Our present setup isolates the contribution of retrieval-grounded prompting while preserving extensibility to multi-strategy systems.

RAILS demonstrates a practical design trade-off, prioritizing semantic fidelity over end-to-end automation, making it a reproducible and verifiable baseline for advancing LLM-powered code repair.

\balance

\section{Conclusion}\label{sec:conclusion}
This paper presented RAILS, a retrieval-augmented framework that enhances Java code synthesis by combining semantic search with a validation feedback loop to improve the reliability of LLM-generated code. Evaluated across 78 Java import error scenarios, including standard APIs, deprecated libraries, GUI toolkits, and custom utilities, RAILS consistently produced accurate and context-based fixes. It succeeded in 100\% of semantically verifiable cases, even when compilation failed due to missing dependencies. In contrast, baseline prompting often hallucinated fixes or altered logic, especially in cases requiring third-party or project-specific knowledge.

The modular architecture of RAILS enables reproducibility, extensibility, and plug-and-play model integration, making it well-suited for modern IDEs. The planned extensions include hybrid symbolic-semantic retrieval (via pgvector), cross-language repair capabilities (e.g., Python, C++), and expanded benchmarks using automated pipelines. In the long term, our goal is to deploy RAILS as a real-time plugin for educational and industrial IDEs, providing verifiable context-aware LLM support for robust developer assistance~\cite{pinecone2024, yuan2402llm}.

\subsection{Ethical Considerations} RAILS sources information from publicly available documentation and community forums. As such, the system may inherit outdated practices, biased examples, or incorrect assumptions from its retrieval corpus. Future work will explore filtering and explainability strategies to align retrieved content with modern coding standards and ensure responsible LLM-assisted development.

\section*{Acknowledgment}
This research is funded by the generous support of Concordia University of Edmonton under the Seed Grant Program. A portion of our experiments were executed on Google Colab, and we thank its compute infrastructure for enabling reproducible, cloud-based testing. We also plan to scale experiments on the Digital Research Alliance (https://www.alliancecan.ca) of Canada's Graham cluster in future iterations.

\bibliographystyle{IEEEtran}
\bibliography{bibliography}
\end{document}